\documentclass{blois}

\usepackage{amsmath}
\usepackage{amssymb}

\def\be{\begin{equation}}
\def\ee{\end{equation}}
\def\bea{\begin{eqnarray}}
\def\eea{\end{eqnarray}}

\newcommand{\Photo}{}

\begin{document}
\vspace*{4cm}
\title{Can supernova remnants accelerate protons up to PeV energies?}

\author{S. GABICI$^1$, D. GAGGERO$^2$, F. ZANDANEL$^2$}

\address{$^1$APC, Univ. Paris Diderot, CNRS/IN2P3, CEA/Irfu, Obs. de Paris, Sorbonne Paris Cit\'e, \\75013 Paris, France\\
$^2$GRAPPA Institute, University of Amsterdam, 1098 XH Amsterdam, The Netherlands
}

\maketitle\abstracts{
Supernova remnants are believed to be the sources of galactic cosmic rays. Within this framework, diffusive shock acceleration must operate in these objects and accelerate protons all the way up to PeV energies. To do so, significant amplification of the magnetic field at the shock is required.
The goal of this paper is to investigate the capability of supernova remnants to accelerate PeV protons.
We present analytic estimates of the maximum energy of accelerated protons under various assumptions about the field amplification at supernova remnant shocks.
We show that acceleration up to PeV energies is problematic in all the scenarios considered.
This implies that either a different (more efficient) mechanism of field amplification operates at supernova remnant shocks, or that the sources of galactic cosmic rays in the PeV energy range should be searched somewhere else.}

\section{Introduction}

The Hillas criterium is very often invoked in order to estimate the maximum energy $E_{max}$ of Cosmic Rays (CRs) that can be accelerated within a given astrophysical object.
For a SuperNova Remnant (SNR) shock of radius $R_s$, shock velocity $u_s$ and downstream magnetic field $B_s$, the Hillas criterium reads \cite{hillas}:
\begin{equation}
\label{eq:hillas}
E_{max} = \epsilon ~\left( \frac{R_s}{\rm pc} \right) \left( \frac{u_s}{1000~{\rm km/s}} \right) \left( \frac{B_s}{\mu {\rm G}} \right) ~ {\rm TeV}
\end{equation}
where $\epsilon$ is a parameter of order unity.
A possible way to estimate the numerical value of $\epsilon$ has been presented, among many others, by Ptuskin and Zirakashvili \cite{ptuskinzirakashvili2005}, and is based on the fact that the diffusion length ahead of a shock of a particle of energy $E$ is of the order of $D/u_s$, where $D = Ec/3 q B_{up}$ is the Bohm diffusion coefficient upstream of the shock ($q$ is the elementary charge). 
When the diffusion length becomes comparable to some fraction (typically 5 to 10\%) of the shock radius, particles are assumed to escape.
From these considerations one can then derive $E_{max}$ and by comparison with Equation \ref{eq:hillas} obtain $\epsilon \sim 1/3$. Other approaches give similar values \cite{hillas,lagagecesarsky}.

The dynamical evolution of a young SNR shock in the ejecta dominated phase can be described by the self-similar expressions derived by Chevalier \cite{chevalier1982}.
For a type Ia supernova, exploding in a uniform insterstellar medium of density $n$, the shock radius and velocity depend on time $t_{\rm kyr} = t/(1000$ yr) as:
\begin{eqnarray}
\label{eq:Ia}
R_s &=& 5.3 \left( \frac{E_{51}^2}{M_{ej,\odot} n} \right)^{1/7} t_{\rm kyr}^{4/7} ~ {\rm pc} \\
u_s &=& 3.0 \times 10^3 \left( \frac{E_{51}^2}{M_{ej,\odot} n} \right)^{1/7} t_{\rm kyr}^{-3/7} ~ {\rm km/s}
\end{eqnarray}
where $E_{51}$ is the explosion energy in units of $10^{51}$ erg and $M_{ej,\odot}$ the mass of the ejecta in solar masses.
For type II supernovae exploding in a circumstellar wind of density $n(R) \propto R^{-2}$ one has:
\begin{eqnarray}
R_s &=& 7.7 \left( \frac{E_{51}^{7/2} u_{w,6}}{\dot{M}_{-5} M_{ej,\odot}^{5/2}} \right)^{1/8} t_{\rm kyr}^{7/8} ~ {\rm pc} \\
u_s &=& 6.6 \times 10^3 \left( \frac{E_{51}^{7/2} u_{w,6}}{\dot{M}_{-5} M_{ej,\odot}^{5/2}} \right)^{1/8} t_{\rm kyr}^{-1/8} ~ {\rm km/s}
\label{eq:II}
\end{eqnarray}
where $u_{w,6}$ is the wind speed in units of 10 km/s and $\dot{M}_{-5}$ is the wind mass loss rate in units of $10^{-5} M_{\odot}$/yr.
The expressions above are valid as long as the mass of the ejecta dominates over the mass of the interstellar medium swept up by the shock.
For type Ia supernovae this is true for times smaller than $\sim 0.17 ~M_{ej,\odot}^{5/2} n^{-1/3} E_{51}^{-1/2}$ kyr, while for type II supernovae the limiting time is $\sim 0.097 ~u_{w,6} M_{ej,\odot}^{3/2} \dot{M}_{-5}^{-1} E_{51}^{-1/2}$ kyr.

After combining Equations \ref{eq:Ia} to \ref{eq:II} with the Hillas criterium, and assuming a typical interstellar magnetic field $B_{ISM} \sim 3 ~\mu$G compressed by a factor of 4 at the shock one gets $E_{max}^{Ia} \sim 0.1 ~ t_{\rm kyr}^{1/7} ~ {\rm PeV}$ for type Ia supernovae
(with $E_{51} = 1$, $M_{ej,\odot} = 1.4$, and $n = 0.1$ cm$^{-3}$) and $E_{max}^{II} \sim 0.07 ~t_{\rm kyr}^{3/4} ~ {\rm PeV}$ for type II supernovae
(with $E_{51} = u_{w,6} = \dot{M}_{-5} = 1$ and $M_{ej,\odot} = 5$).
In both cases, the maximum energy is much smaller than a PeV. On the other hand, acceleration up to at least $\approx 1$ PeV is needed in order to explain the observed spectrum of CR protons, which extends up to such energies \cite{kascade,argo}.

In order to overcome this problem, several mechanism for magnetic field amplification at shocks have been proposed. An amplification of a factor well above 10 is needed in order to reach energies of few PeVs. Remarkably, X-ray observations revealed that such fields are indeed present  downstream of the shock of many SNRs \cite{jacco}. 
Bell \cite{bell04} proposed that the magnetic field upstream of the shock might be largely amplified by a non-resonant instability driven by the stream of CRs escaping the shock. 
Another possibility is to amplify the upstream field as the result of turbulence induced by the CR gradient upstream of the shock acting on an inhomogeneous ambient medium. This mechanism is often referred to as Drury instability~\cite{turlog}.
Finally, Giacalone and Jokipii \cite{giacalone} suggested that preexisting turbulent density fluctuations upstream of the shock might induce significant amplification of the downstream magnetic field.
The goal of this paper is to discuss critically all these possibilities in order to identify the conditions (if any) under which the magnetic field amplification induces acceleration of protons up to the PeV energy domain.

\section{Bell instability}

CRs escaping a shock carry a current $j_{CR}$. 
This is balanced by an oppositely directed current in the background plasma.
The resulting ${\bf j}_{CR} \times {\bf B}$ force is responsible for the growth of initial small perturbations in the field.
Such instability is non-resonant, in the sense that initial small perturbations in the magnetic field grow the most rapidly on a scale much smaller than the CR Larmor radius, at a rate \cite{bell04,martina}:
\begin{equation}
\label{eq:growth}
\gamma_{max} \sim \frac{j_{CR}}{c} \sqrt{\frac{4 \pi}{\varrho}}
\end{equation}
where $j_{CR}$ is the current carried by the CRs escaping the remnant and $\varrho$ is the density of the upstream medium (assumed here to be fully ionised).
The current $j_{CR}$ is carried by the highest energy CRs, which are non-magnetised and are the only ones able to escape the shock. 
Within the framework of diffusive shock acceleration theory, the current $j_{CR}$ at the position of the shock front is given by \cite{bell2013}:
\begin{equation}
\label{eq:current}
j_{CR} = e u_s \pi p_{max}^3 f_0(p_{max})
\end{equation}
where $f_0 \propto p^{-\alpha}$ is the differential energy distribution of CRs at the shock, and $E_{max} = p_{max} c$ is the maximum energy of CRs.
It is convenient to express the CR current in terms of the total CR pressure at the shock:
\begin{equation}
P_{CR} = \frac{4 \pi}{3} \int {\rm d} p p^2 f_0(p) p c
\end{equation}
which gives:
\begin{equation}
\label{eq:current}
j_{CR} = \frac{e u_s P_{CR}}{c p_{max}} g(\alpha,p_{max})
\end{equation}
where:
\begin{equation}
g(\alpha,p_{max}) = \frac{3}{4} (\alpha - 4) \left[ \left( \frac{p_0}{p_{max}} \right)^{4-\alpha} - 1 \right]^{-1}
\end{equation}
for $\alpha \ne 4$ and
\begin{equation}
g(\alpha,p_{max}) = \frac{3}{4} \left[ \ln(\frac{p_{max}}{p_0}) \right]^{-1}
\end{equation}
for $\alpha = 4$. In both expressions $p_0 c \sim 1$ GeV.
Note that for $\alpha \approx 4$, as expected for diffusive shock acceleration, $g$ is a slowly varying function of the maximum momentum $p_{max}$.
For convenience, in the following we will consider the values of $g$ at a specific energy $E_*$, $\tilde{g}(\alpha) = g(\alpha,E_*)$. The values of $\tilde{g}$ for $E_{*} = 100$ TeV ($E_{*} = 1$ PeV), are equal to 0.11 (0.10), 0.065 (0.054), 0.035 (0.025), and 0.017 (0.010) for $\alpha = $ 3.9, 4.0, 4.1, and 4.2, respectively.

At a given position $r$ upstream of the shock and at a given time $t$, the maximum level of field amplification is determined by the parameter $a = \int {\rm d}t ~ \gamma_{max}$, representing the number of e-foldings of the field (i.e.~a small initial perturbation is amplified by a factor of $e^a$).
Given that the highest energy CRs are assumed to free stream away from the shock, the expression for $j_{CR}$ given in Equation \ref{eq:current} has to be multiplied by a geometrical factor $(R_s/r)^2$ that represents the current at the position $r$ carried by CRs escaped when the shock had a radius $R_s$. 
From the definition of the parameter $a$, and after making use of Equations \ref{eq:growth} and \ref{eq:current} one can obtain the following integral equation \cite{bell2013,klara1,klara2}:
\begin{equation}
\left( \frac{c a}{e} \right) r^2 \sqrt{\frac{\varrho}{4 \pi}} = \int_0^r {\rm d}R_s \frac{P_{CR} R_s^2}{c p_{max}} \tilde{g}(\alpha)
\end{equation}
which after being differentiated with respect to $r$ gives the value of the maximum energy of CRs:
\begin{equation}
\label{eq:Emaxbell}
E_{max} = \frac{4 \sqrt{\pi} e \eta}{a c (4-s)} \tilde{g}(\alpha) \varrho^{1/2} u_s^2 R_s
\end{equation}
where $s = 0$ for a uniform interstellar medium medium, and $s = 2$ for a circumstellar medium with a density profile $\varrho \propto r^{-2}$, and where we introduced a CR acceleration efficiency $\eta$ defined as the fraction of the shock ram pressure converted into CR pressure:
$P_{CR} = \eta \varrho u_s^2$.
Equation \ref{eq:Emaxbell} suffices to estimate the value of the maximum energy of accelerated CRs.
However, it can be recast in a more familiar way by introducing another efficiency parameter $\xi_B$, defined as:
$B_s^2/8 \pi = \xi_B \varrho u_s^2$,
representing the fraction of the shock ram pressure converted into downstream magnetic pressure.
Equation \ref{eq:Emaxbell} then reads:
\begin{equation}
\label{eq:likeHillas}
E_{max} = \frac{\sqrt{2} e ~ \eta ~ \xi_B^{-1/2}}{a c (4-s)}~ \tilde{g}(\alpha)~ B_s ~u_s R_s
\end{equation}
which is equivalent to the Hillas criterium.

We are now in the position to estimate the maximum energy of accelerated protons during the ejecta dominated phase of the SNR evolution. After combining Equations \ref{eq:Ia} to \ref{eq:II} with Equation \ref{eq:Emaxbell} we get:
\begin{equation}
\label{eq:EmaxIa}
\frac{E_{max}}{\rm PeV} \sim 0.3 \left( \frac{\eta}{0.3} \right) \left( \frac{5}{a} \right) \left( \frac{\tilde{g}}{0.065} \right) \left( \frac{M_{ej,\odot}}{1.4} \right)^{-\frac{3}{7}} E_{51}^{\frac{6}{7}} n^{\frac{1}{14}} \left( \frac{t}{100~{\rm yr}} \right)^{-\frac{2}{7}}  
\end{equation}
for type Ia supernovae and
\begin{equation}
\label{eq:EmaxII}
\frac{E_{max}}{\rm PeV} \sim 0.3 \left( \frac{\eta}{0.3} \right) \left( \frac{5}{a} \right) \left( \frac{\tilde{g}}{0.065} \right)  E_{51}^{\frac{7}{8}} \left( \frac{\dot{M}_{-5}}{u_{w,6}} \right)^{\frac{1}{4}} \left( \frac{M_{ej,\odot}}{5} \right)^{-\frac{5}{8}} \left( \frac{t}{100~{\rm yr}} \right)^{-\frac{1}{4}} 
\end{equation}
for type II supernovae. 
All the parameters on the right hand side of the equations above have been normalized as in Schure and Bell \cite{klara1}, and a = 5 is the number of e-foldings of the field when the mechanism of amplification is found to saturate in numerical simulations \cite{bell2013}.
{\it We found then that in this scenario the acceleration up to PeV energies is possible only in the very early phase of the SNR evolution, and only if the spectrum of accelerated particles is $\alpha = 4$ or harder}. For softer spectra the parameter $\tilde{g}$ decreases significantly making the acceleration of PeV protons very challenging. This is in agreement with the findings of Schure and Bell \cite{klara1} (though they considered an almost constant velocity for the initial phase of type Ia SNRs, and consequently obtained a lower value for $E_{max}$) and of Cardillo et al. \cite{martina} (where only type II SNR were considerd).

For the parameter normalisations adopted in Equation \ref{eq:EmaxIa} and \ref{eq:EmaxII} a maximum energy of 1~PeV corresponds to a SNR age much smaller than a century for both type Ia and type II supernovae.
Thus, the question arises whether in such a short time SNRs can convert a sufficient amount of energy
into PeV protons, and explain the observed intensity of CRs in the region of the knee. 
In fact, it is known that in order to reproduce the observed CR spectrum, particles at the knee should be accelerated at SNR shocks at the transition between the ejecta dominated and the Sedov phase \cite{ptuskinzirakashvili2005}, which happens when the SNR has a typical age of few centuries. 
By imposing that a maximum energy of $\sim 1$ PeV is indeed reached at such transition, and after adopting $\alpha = 4$ and $a = 5$, one gets that the two quantities:
\begin{equation}
\left( \frac{\eta}{0.3} \right) \left( \frac{M_{ej,\odot}}{1.4} \right)^{-8/7} E_{51}~ n^{1/6} ~~~~ {\rm and} ~~~~
\left( \frac{\eta}{0.3} \right) \left( \frac{M_{ej,\odot}}{5} \right)^{-1} \left( \frac{\dot{M}_{-5}}{u_{w,6}} \right)^{1/2} E_{51} 
\end{equation}
have to be of the order of $\lesssim 10$.
This condition is difficult to be achieved for type Ia supernovae, but might be satisfied for certain choices of parameters values for type II supernovae (e.g. a low value for $M_{ej,\odot}$ and a slow and dense wind~\cite{martina}).
Unfortunately, the assumption $\alpha = 4$ made above in order to facilitate the acceleration up to PeV energies is in some tension with studies of the CR propagation in the galaxy, which rather indicate $\alpha > 4$ \cite{strongreview,fermidiff}.
As we have seen above, for such steep spectra the acceleration of PeV particles is a real challenge.

The scenario proposed by Bell is, at the moment, virtually untestable in a direct way, given that we don't know any SNR in the Galaxy younger than a century \cite{reynolds}.
Additionally, the constraints coming from propagation studies in the Galaxy refer to particle energies well below the PeV.
However, even though this would require quite some fine tuning, one might speculate that (non observed) very young SNRs do accelerate CRs with a spectrum harder than that produced at later ages.

\section{Drury instability}

This is an acoustic instability first presented by Drury and Falle \cite{falle}, and operates upstream of shocks, where a spatial gradient in the CR pressure $\nabla P_{CR}$ exists.
The pressure gradient exerts a force onto the thermal plasma, which induces an acceleration on a fluid element of density $\varrho$ equal to $-\nabla P_{CR}/\varrho$.
If the upstream medium is clumpy, with density inhomogeneities of amplitude $\delta \varrho/\varrho$, then fluctuations in the acceleration field will also appear. These will in turn induce velocity fluctuations and therefore further density fluctuations \cite{turlog,jones,turlogbis,delvalle}.

Drury and Downes \cite{turlog} considered a plane shock located at a position $z = L$, and a linearly rising CR pressure profile in the precursor:
\begin{equation}
\label{eq:linear}
P_{CR}(z) = \eta \varrho u_s^2 \frac{z}{L}
\end{equation} 
where $\eta$ is the same CR acceleration efficiency defined in Equation \ref{eq:Emaxbell}.
Note that $P_{CR}(L) = 0$ so $L$ represents the length of the CR precursor.
Under these circumstances, the force exerted onto the upstream fluid is spatially homogeneous and equal to $-\eta \varrho u_s^2/L$.
The force will act on a perturbation $\delta \varrho$ for an advection time $L/u_s$, producing a velocity fluctuation:
\begin{equation}
\delta u \sim \frac{\delta \varrho}{\varrho} \eta u_s ~~ .
\end{equation}
It follows that the total kinetic energy in the turbulence is of the order of:
\begin{equation}
\epsilon_T \sim \frac{1}{2} \varrho (\delta u)^2 \sim \left( \frac{\delta \varrho}{\varrho} \right)^2 \frac{\eta^2 \varrho u_s^2}{2}
\end{equation}
After setting this equal to $B^2/8 \pi$ one sees that the maximum possible value for the amplified (upstream) magnetic field in this scenario is:
\begin{equation}
B \sim \sqrt{4 \pi}~ \eta \frac{\delta \varrho}{\varrho} \left( \varrho u_s^2 \right)^{1/2}
\end{equation}
Drury and Downes \cite{turlog} then suggested that if the instability saturates when $\delta \varrho/\varrho \sim 1$ and if CR acceleration is very efficient ($\eta \sim 1$) then the magnetic field can be significantly amplified.

One advantage of this scenario is that, while the Bell instability excites turbulence over spatial scales much smaller (i.e. non-resonating) than the particles' Larmor radii, here the injection length scale $\lambda$ of magnetic turbulence is most likely much larger, being  the same as that of density fluctuations. 
In order to drive a cascade, the turbulence turnover time $\lambda/\delta u$ has to be much smaller than the time available for the instability to operate $L/u_s$:
\begin{equation}
\label{eq:constrain}
\lambda \ll \eta \frac{\delta \varrho}{\varrho} L
\end{equation}
which is easily satisfied.

Indeed, the linear profile in the CR pressure (Equation \ref{eq:linear}) is justified only in the extreme situation characterised by $\eta \sim 1$ and by a very large gas compression in the shock precursor, induced by the presence of CRs. In this case the spectrum of the accelerated particles is very hard: $f_0 \propto p^{-3.5}$, as first shown by Malkov \cite{malkov}.
In fact, various heating processes operating in the shock precursor and the dynamical effect of an amplified field at the shock transition are likely to limit the shock compression to values quite close to the standard value $r \gtrsim 4$. This also results in spectra of accelerated particles which are quite close to power laws $f_0 \propto p^{-\alpha}$ with $\alpha$ larger than, but quite close to, 4 \cite{drift,damiano} (see also the review by Blasi \cite{pasquale}).
Under these circumstances the CR spatial distribution in the shock precursor can be described by the expression:
\begin{equation}
\label{eq:realistic}
f(p,z) = f_0(p) \exp \left[ \frac{(z-L) ~u_s}{D(p)}  \right] \approx f_0(p)~ \vartheta \left[ L-\frac{D(p)}{u_s} \right] 
\end{equation}
where $f_0(p) \propto p^{-\alpha}$ and $\vartheta[...]$ is the Heaviside function.
The second approximate equality was first introduced by Eichler \cite{eichler} and states that no particles of momentum $p$ can reach distances upstream of the shock larger than the diffusion length $D(p)/u_s \equiv L-z$.
From the definition of the precursor length $L = D(p_{max})/u_s$ and from the assumption of (spatially homogeneous) Bohm diffusion, one gets $p/p_{max} = 1 - (z/L)$.
The CR pressure profile in the precursor can then be computed as:
\begin{equation}
P_{CR}(z) \approx \frac{4 \pi}{3} \int_{p(z)}^{p_{max}} {\rm d}\hat{p} ~\hat{p}^2 f_0(\hat{p}) ~\hat{p} c
\end{equation}
It can be seen that in this case the CR pressure drops to $1/e$ times its value at $z = L$ at a distance from the shock of $\epsilon = 1-z/L \sim 5 \times 10^{-3}$ and $\sim 10^{-3}$ for $\alpha = 4$ and $4.1$, respectively, and $p_{max}c = 100$ TeV.
These are very small distances, compared to that obtained from Equation \ref{eq:linear}, which gives $\epsilon = 1-z/L \sim 0.6$.

Given these numbers, the question arises about the actual effectiveness of the amplification mechanism. The magnetic field amplification operates over a region ahead of the shock defined by the spatial scale of the gradient of the CR pressure, which can be expressed as $\epsilon L$. For a linear profile of the CR pressure (Equation \ref{eq:linear}) $\epsilon$ is of order unity, and Equation \ref{eq:constrain} applies. However, for the more realistic situation described by Equation \ref{eq:realistic} one has $\epsilon \ll 1$. The right hand side of Equation \ref{eq:constrain} should then be multiplied by $\epsilon$, making the condition difficult to be satisfied, and casting doubts on the viability of the field amplification.
This suggests that {\it this mechanism could lead to the production of PeV CRs only in the extreme case of a strongly modified shock ($\eta \sim 1$) characterised by a large compression factor in the shock precursor, and a very hard spectrum of accelerated CRs (asimptotically $\propto p^{-3.5}$) \cite{malkov}}. These conditions would result in a linear profile of the upstream CR pressure as the one in Equation \ref{eq:linear}.

\section{Giacalone \& Jokipii's approach}

Finally, an alternative way to amplify the magnetic field at SNR shocks has been proposed by Giacalone and Jokipii \cite{giacalone}, and then discussed in several papers \cite{balsara,inoue,inoue2,guo,ji}. This mechanism is not related to the presence of CRs at the shock, but rather to the dynamics of magnetised fluids.
Also in this scenario, as in the Drury instability, density fluctuations in the interstellar medium play a crucial role.
Density  inhomogeneities in the background fluid are expected to warp the shock surface, and this in turn generates vorticity in the downstream fluid flow, resulting in an amplification of the field.
This mechanism operates downstream of the shock only, and can work in combination with the mechanisms described above, which instead amplify the upstream magnetic field.
Thus, it seems that {\it this mechanism alone cannot be responsible for the increase in the CR acceleration rate needed to explain PeV CRs}, because particles of very high energy would not be confined upstream of the shock. 

\section{Critical discussion and conclusions}

In this paper we described briefly various mechanisms for the amplification of the magnetic field at SNR shocks, in order to assess their capability of boosting the acceleration of CRs up to PeV energies. 
The non-resonant, current-driven instability proposed by Bell \cite{bell04,bell2013} is by far the most popular mechanism and received a lot of attention (see e.g. the recent review \cite{klarareview} and the long list of references therein).
Even though under some circumstances this mechanism might indeed induce the acceleration of PeV protons at SNR shocks, it is believed to do so for a very short time only (definitely less than a century).
This makes the scenario very difficult to be tested (at the estimated rate of 3 supernovae per century in the Galaxy, we expect to have at most $\approx 1$ SNR currently accelerating PeV particles), and poses some doubts about the capability of SNRs to accelerate enough particles to PeV energies and match the observed intensity of the CR spectrum in the region of the knee. 
Moreover, the acceleration to PeV energies requires hard spectra ($\propto p^{-4}$ or harder), in tension with constraints from studies of CR propagation which point towards spectra definitely softer than $p^{-4}$.

The amplification mechanism resulting from the Drury instability \cite{turlog} would definitely deserve equal attention because it might amplify the field to values even larger than those predicted by the Bell instability. However, also in this case it seems that quite extreme conditions are required, such as an extremely large CR acceleration efficiency ($\eta \sim 1$) and a very hard spectrum of the accelerated particles. 
Interestingly, modifications of this scenario have been presented (see e.g. \cite{malkovdiamond} and references therein), and might provide an additional boost to the CR acceleration rate. 

Unfortunately, no direct observational evidence for the acceleration of PeV protons at SNR shocks has ever been reported. The observational difficulties are mainly related to the fact that probably very few SNRs (if any) are currently accelerating particles up to the PeV domain, and also to the fact that the most distinctive signatures of the presence of PeV protons are multi-TeV neutrinos and multi-TeV gamma-rays produced in the inelastic interactions between CRs and ambient gas \cite{felixbook}.
The sensitivity of neutrino telescopes is still not at the level needed to detect the expected signal from SNRs, and the detection of multi-TeV (up to hundreds of TeV) photons is a task at the limit of the capabilities of current instruments.
Despite this, the search for direct or indirect observational signatures of acceleration of PeV protons at SNRs remains a priority (see e.g. \cite{me2007}).

Remarkably, the H.E.S.S. Collaboration recently reported on the first detection of a Galactic PeVatron, located in the centre of the Galaxy and most likely related to the central supermassive black hole \cite{pevatron}.
This discovery opens new perspectives, since it demonstrates the existence of galactic particle PeVatrons other than SNRs, and reminds us that scenarios alternative to the SNR paradigm for the origin of CRs should not be discarded.

\section*{Acknowledgments}

SG acknowledges support from the Observatory of Paris under the programme {\it Action F\'ed\'eratrice CTA}. He also thanks Luke Dury and Giovanni Morlino for useful discussions.
FZ acknowledges the support by the Netherlands Organization for Scientific Research (NWO) through a Veni grant.

\end{document}